\documentclass[aps,prd,twocolumn,groupedaddress,showpacs]{revtex4}

\usepackage{graphicx,dcolumn,bm,amssymb,amsmath,latexsym,amsfonts}

\begin{document}

%%%%%%%%%%%%%%%%%%%%%%%%%%%%%%%%%%%%%
%defining some commands
\newcommand{\nonu}{\nonumber}
\newcommand{\sm}{\small}
\newcommand{\noi}{\noindent}
\newcommand{\npg}{\newpage}
\newcommand{\nl}{\newline}
\newcommand{\bp}{\begin{picture}}
\newcommand{\ep}{\end{picture}}
\newcommand{\bc}{\begin{center}}
\newcommand{\ec}{\end{center}}
\newcommand{\be}{\begin{equation}}
\newcommand{\ee}{\end{equation}}
\newcommand{\beal}{\begin{align}}
\newcommand{\eeal}{\end{align}}
\newcommand{\bea}{\begin{eqnarray}}
\newcommand{\eea}{\end{eqnarray}}
\newcommand{\bnabla}{\mbox{\boldmath $\nabla$}}
\newcommand{\univec}{\textbf{a}}
\newcommand{\VectorA}{\textbf{A}}
\newcommand{\Pint}
%%%%%%%%%%%%%%%%%%%%%%%%%%%%%%%%%%%%

\title{Unequal binary configurations of interacting Kerr black holes}

\author{I. Cabrera-Munguia\footnote{icabreramunguia@gmail.com}}
\affiliation{ Departamento de F\'isica y Matem\'aticas, Universidad Aut\'onoma de Ciudad Ju\'arez, 32310 Ciudad Ju\'arez, Chihuahua, M\'exico}
%\date{\today}

%------------------------------Begin of the document --------------------------------

\begin{abstract}
Stationary axisymmetric binary configurations of unequal Kerr sources with a massless strut among them are developed in a physical representation. In order to describe interacting black holes, the axis conditions in the most general case are solved analytically deriving the corresponding $5$-parametric asymptotically flat exact solution. In addition, we obtain concise formulas for the black hole horizons, the interaction force, as well as the thermodynamical characteristics of each source in terms of physical Komar parameters: mass $M_{i}$, angular momentum $J_{i}$, and coordinate distance $R$, where such parameters are contained inside of the coefficients of a cubic equation which can be interpreted as a dynamical law for interacting black holes with struts. Some limits are obtained and discussed.
\end{abstract}

\maketitle

\section{Introduction}
\vspace{-0.5cm}
Nowadays, the coalescence process among two interacting black hole (BH) sources has been considered an outstanding candidate to study and detect gravitational waves (GW) by the LIGO and Virgo Collaborations \cite{Ligo}. Due mainly to the fact that the numerical simulations are the main tool at the moment of considering the merging process (MP) between two BHs, it motivates scientists around the world to match this huge discovery with models within the framework of exact solutions. Nevertheless, it seems quite complicated to construct exact results describing physical models that can take into account all the possible interactions between the components of the binary system (BS) during the MP. In this respect, the double-Kerr-NUT (DKN) solution \cite{KramerNeugebauer} developed by Kramer and Neugebauer almost four decades ago permits us to describe a realistic interaction between to massive rotating sources in stationary axisymmetric spacetimes with some issues regarding the regularity of the solution, since it is well-known that in the absence of a supporting strut (conical singularity \cite{Bach,Israel}) ring singularities off the axis appear if at least one of the masses turns out to be negative \cite{Hoenselaers, MRS,Hennig}, even yet if the positive mass theorem \cite{SchoenYau1, SchoenYau2} is fulfilled in the BS. The last point suggests us to focus our attention in configurations of unequal binary BHs with a conical singularity in between, with the main purpose of describing their dynamical and physical properties before the coalescence process may occur. Nevertheless, until this day solving analytically the axis conditions in the most general case has been one of the main technical (highly complicated) problems to treat binary configurations of interacting BHs.

The present paper aims at solving for the first time the axis conditions in order to derive a $5$-parametric subclass of the DKN solution \cite{KramerNeugebauer}. Until now we had thought that reaching such a goal was almost unthinkable. The solution represents the most general case with regards to the description of the dynamical interaction of binary configurations of unequal Kerr sources, with the main distinctive of being characterized by five arbitrary physical Komar parameters \cite{Komar}: the masses $M_{i}$ and angular momenta $J_{i}$, as well as the coordinate distance $R$. These parameters are part of a dynamical law for interacting BHs, which in the absence of a supporting strut becomes in the equilibrium law for two nonequal Kerr constituents \cite{MR}. A remarkable feature in our analysis of the interaction force related to the conical singularity reveals the existence of equilibrium states without strut during the MP where both BHs are endowed with positive masses.

The outline of the paper is as follows. In Sec. II we establish a physical representation of the DKN problem as a $7$-parametric non-asymptotically flat exact solution. This will lead us to consider the suitable parametrization that allows to kill first the NUT charge \cite{NUT}, and later on, to solve the axis condition that disconnects the middle region among sources. In Sec. III we obtain concise formulas for both event horizons as a function of physical Komar parameters, with the main objective to determine some dynamical and thermodynamical characteristics of the BS. Final remarks are presented in Sec. IV.

\vspace{-0.6cm}
\section{The double-Kerr-NUT solution in a physical representation}
\vspace{-0.5cm}
The well-known DKN solution constructed by Kramer and Neugebauer long time ago \cite{KramerNeugebauer} represents a superposition of two massive rotating sources in stationary spacetimes. It was developed by employing B\"{a}cklund transformations \cite{Neugebauer} as a modern generation technique of exact solutions in Einstein's vacuum equations. Moreover, the DKN solution can also be derived through the Sibgatullin method (SM) \cite{Sibgatullin} which is also very fit to describe electrovacuum spacetimes \cite{RMJ}. Both approaches start with a particular form of the Ernst potential \cite{Ernst} on the symmetry axis (the axis data), which is then extended in the whole spacetime. According to Ernst formalism \cite{Ernst}, the vacuum Einstein field equations are reduced into a new complex equation for solving
\vspace{-0.2cm}
\be ({\cal{E}}+ \bar{\cal{E}})({\cal{E}}_{\rho \rho} + \rho^{-1}{\cal{E}}_{\rho}+{\cal{E}}_{z z} )=2({\cal{E}}_{\rho}^{2}+{\cal{E}}_{z}^{2}), \label{Ernsteq}\ee

\noi being ${\cal{E}}$ defined in Weyl-Papapetrou cylindrical coordinates $(\rho,z)$, where the subscript $\rho$ or $z$ denotes partial differentiation. In this regard, the line element for stationary axisymmetric spacetimes is given by \cite{Papapetrou}
\be ds^{2}=f^{-1}\left[e^{2\gamma}(d\rho^{2}+dz^{2})+\rho^{2}d\varphi^{2}\right]- f(dt-\omega d\varphi)^{2},
\label{Papapetrou}\ee

\noi where the metric functions $f(\rho,z)$, $\omega(\rho,z)$ and $\gamma(\rho,z)$ can be derived from the following system of differential equations:
\bea \begin{split}  f&=  {\rm{Re}}({\cal{E}}), \\
\omega_{\rho} &= -4\rho ({\cal{E}}+ \bar{\cal{E}})^{-2}{\rm{Im}}({\cal{E}}_{z}) ,\\
\omega_{z} &= 4\rho ({\cal{E}}+ \bar{\cal{E}})^{-2}{\rm{Im}}({\cal{E}}_{\rho}) ,\\
\gamma_{\rho}&=\rho ({\cal{E}}+ \bar{\cal{E}})^{-2} \left({\cal{E}}_{\rho} \bar {\cal{E}}_{\rho} -{\cal{E}}_{z} \bar {\cal{E}}_{z}\right),\\
\gamma_{z}&=2\rho ({\cal{E}}+ \bar{\cal{E}})^{-2} \rm{Re}({\cal{E}}_{\rho}\,{\bar{\cal{E}}}_{z}),
\label{metrics}\end{split}\eea

\noi once we know a peculiar form of the Ernst potential ${\cal{E}}$. For solving the nonlinear Eq.\ (\ref{Ernsteq}) by using the SM, the axis data for the Ernst potential in vacuum systems adopts the most general representation as follows \cite{RMJ}:
\be {\cal{E}}(\rho=0,z):=e(z)=1+\sum_{i=1}^{2}\frac{e_{i}}{z-\beta_{i}},\label{generalernst}\ee

\noi where $\{e_{i},\, \beta_{i}\}$, $i=1,2,$ are arbitrary complex constants related to the Geroch-Hansen (GH) multipole moments \cite{Geroch,Hansen}. At the same time, the SM begins with the characteristic equation
\vspace{-0.2cm}
\be e(z) + \bar{e}(z)=0,\label{characteristic}\ee

\noi being $\alpha_{n}$, for $n=\overline{1,4}$, the roots of Eq.\ ({\ref{characteristic}}) that locate the sources on the symmetry axis. In order to change the old parameters $\{e_{i},\beta_{i}\}$ by the new ones $\{\alpha_{n},\beta_{i}\}$,  Eq.\ (\ref{generalernst}) is placed into Eq.\ (\ref{characteristic})
\vspace{-0.2cm}
\be 2+ \sum_{i=1}^{2}\left(\frac{e_{i}}{z-\beta_{i}}+ \frac{\bar{e}_{i}}{z-\bar{\beta}_{i}}\right)
= \frac{2\prod_{n=1}^{4}(z-\alpha_{n})}{\prod_{i=1}^{2}(z-\beta_{i})
(z-\bar{\beta}_{i})},
\label{characteristicII} \ee

\noi to obtain
\vspace{-0.2cm}
\bea \begin{split} e_{1}&=\frac{2 \prod_{n=1}^{4}(\beta_{1}-\alpha_{n})}{(\beta_{1}-\beta_{2})(\beta_{1}-\bar{\beta}_{1})
(\beta_{1}-\bar{\beta}_{2})},\\
e_{2}&=\frac{2 \prod_{n=1}^{4}(\beta_{2}-\alpha_{n})}{(\beta_{2}-\beta_{1})
(\beta_{2}-\bar{\beta}_{1})(\beta_{2}-\bar{\beta}_{2})}.
\label{thees}\end{split}\eea

The DKN solution can be performed directly from the last formulas of \cite{RMJ}, with $N=2$ and taking into account an absence of the electromagnetic field $(\Phi=0)$, where such a metric contains eight arbitrary real parameters. However, although the SM allows us to build the DKN solution in all the spacetime, the solution itself lacks of physical meaning at the moment one wishes to study the dynamical interaction between two rotating sources, therefore it is mandatory to solve the axis conditions. At this point, it is worthwhile to stress the fact that solving analytically these conditions cannot be assumed as a trivial problem, and for that reason only identical cases taking the advantage of their symmetry property on the equatorial plane have been considered until this work \cite{Varzugin,MRRS,Costa,CCLP}. Before solving the axis conditions, we are going to depict the DKN problem with an aspect which contains a more physical representation. To do this, we notice that the Fodor-Hoenselaers-Perj\'es procedure \cite{FHP} permits us to calculate from Eq.\ (\ref{generalernst}) the first GH multipolar terms like the total mass of the system $M$, NUT charge $J_{0}$ \cite{NUT}, and total angular momentum $J$ of the system, which are given by
\vspace{-0.2cm}
\be -\frac{e_{1}+e_{2}}{2}=M + i J_{0}, \qquad -\frac{\rm {Im} [e_{1}\beta_{1}+e_{2}\beta_{2}]}{2}=J. \label{total}\ee

\vspace{-0.3cm}
\begin{figure}[ht]
\centering
\includegraphics[width=6.0cm,height=5.0cm]{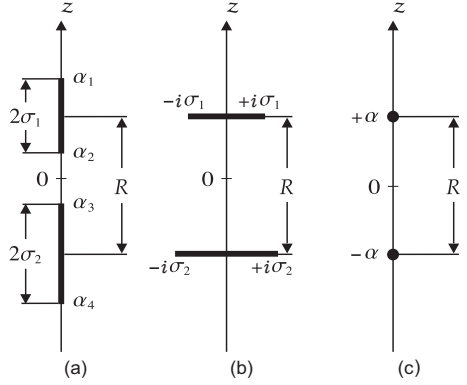}
\caption{Schematic representation of different types of unequal Kerr sources: (a) BH configuration $\sigma_{i}^{2}>0$; (b) hyperextreme sources if $\sigma_{i} \rightarrow i \sigma_{i}$ (or $\sigma_{i}^{2}<0$ ); (c) the extreme limit case $\sigma_{i}=0$.}
\label{DK}\end{figure}

Replacing Eq.\ (\ref{thees}) into Eq.\ (\ref{total}) yields the relation for the total mass
\be \beta_{1}+ \beta_{2}+ \bar{\beta}_{1}+\bar{\beta}_{2}+\sum_{n=1}^{4}\alpha_{n}=-2M \label{themass}\ee

\noi where the parameters $\alpha_{n}$ can be rewritten in terms of the relative distance $R$ and the half-length rod $\sigma_{i}$ as follows:
\bea \begin{split} \alpha_{1}&=\frac{R}{2}+\sigma_{1}, \qquad \alpha_{2}=\frac{R}{2}-\sigma_{1},\\
\alpha_{3}&=-\frac{R}{2}+\sigma_{2}, \qquad \alpha_{4}=-\frac{R}{2}-\sigma_{2}.
\label{thealphas}\end{split}\eea

\noi Thereby we have reduced only one parameter of the DKN solution. It is important to mention that $\sigma_{i}$ can take real positive or pure imaginary values representing BHs (subextreme sources) or relativistic disks (hyperextreme sources), respectively, as shown in Fig.\ \ref{DK}. In addition, to solve Eq.\ (\ref{themass}) one might choose the ansatz for $\beta_{i}$
\be \beta_{1,2}=\frac{-M+i\mathfrak{q} \pm \sqrt{p+i \delta}}{2},\ee

\noi where after using the following parametrization:
\bea \begin{split} p&=R^{2}-\Delta +2\left(\epsilon_{1}-\frac{\epsilon_{2}R}{M}\right)+\frac{2\mathfrak{q}(P_{1}+P_{2})}{M},\\
\delta&=-2(2P_{2}+M\mathfrak{q}),\\
\epsilon_{1,2}&:= \sigma_{1}^{2} \pm \sigma_{2}^{2}, \qquad \Delta:=M^{2}-\mathfrak{q}^{2},
\label{thepdelta}\end{split}\eea

\noi it guides us to simple expressions for the NUT charge and total angular momentum
\begin{widetext}
\bea \begin{split}
J_{0}&= \frac{\mathfrak{q}}{2M} \left( \frac{[\mathfrak{q}(P_{1}+P_{2})-\epsilon_{2}R]^{2}-
M^{2}\left[4P_{1}P_{2}+
(R^{2}-\Delta)(2\epsilon_{1}-\Delta)+\epsilon_{2}^{2}\right]}{\mathfrak{q}^{2}\left[M(R^{2}+M^{2}-
2\Delta)+2\left(\mathfrak{q}(P_{1}+P_{2})+M\epsilon_{1}-R\epsilon_{2}\right)\right]-M(2P_{2}+M \mathfrak{q})^{2}} \right),\\
J&=M \mathfrak{q}-\frac{P_{1}-P_{2}}{2}+\frac{J_{0}P_{2}}{\mathfrak{q}},
\label{Multipolarterms}\end{split}\eea

\noi and now the axis data of the Ernst potential given by Eq.\ (\ref{generalernst}) and satisfying Eq.\ (\ref{characteristic}) results to be
\bea \begin{split}
{\cal E}(0,z)&=\frac{e_{1}}{e_{2}}, \\
e_{1}&=z^{2}-[M + i(\mathfrak{q}+2J_{0})]z + \frac{M(2\Delta-R^{2})-2(M\epsilon_{1}+R\epsilon_{2})}{4M} + \frac{\mathfrak{q}(P_{1} +P_{2})}{2M} -2 \mathfrak{q} J_{0}+ i\left(P_{1}- \frac{2J_{0}(P_{2}+M \mathfrak{q})}{\mathfrak{q}} \right),& \\
e_{2}&=z^{2} + (M -i\mathfrak{q})z + \frac{M(2\Delta-R^{2})-2(M\epsilon_{1}-R\epsilon_{2})}{4M} -\frac{\mathfrak{q}(P_{1} +P_{2})}{2M} + i P_{2}. \label{ernstaxis}\end{split}\eea

The Ernst potential and full metric in the entire spacetime can be reduced eventually until get the
following concise form \cite{MankoRuiz}:
\bea \begin{split}
{\cal{E}}&=\frac{\Lambda+\Gamma}{\Lambda-\Gamma},\qquad
f=\frac{\Lambda \bar{\Lambda}-\Gamma \bar{\Gamma}}{(\Lambda-\Gamma)(\bar{\Lambda}-\bar{\Gamma})}, \qquad \omega=2\mathfrak{q} +\frac{2{\rm{Im}}\left[(\Lambda-\Gamma)\bar{\mathcal{G}}\right]}{\Lambda \bar{\Lambda}-\Gamma \bar{\Gamma}},\qquad e^{2\gamma}=\frac{\Lambda \bar{\Lambda}-\Gamma \bar{\Gamma}}{\kappa_{o} \bar{\kappa}_{o} r_{1}r_{2}r_{3}r_{4}},\\
\Lambda&=4\sigma_{1}\sigma_{2}(\mathfrak{r}_{1}-\mathfrak{r}_{3})(\mathfrak{r}_{2}-\mathfrak{r}_{4})
-[R^{2}-(\sigma_{1}-\sigma_{2})^{2}](\mathfrak{r}_{1}-\mathfrak{r}_{2})(\mathfrak{r}_{3}-
\mathfrak{r}_{4}),\\
\Gamma&=2\sigma_{2} (R^{2}+\epsilon_{2})(\mathfrak{r}_{1}-\mathfrak{r}_{2})+2\sigma_{1}(R^{2}-\epsilon_{2})
(\mathfrak{r}_{3}-\mathfrak{r}_{4})-4\sigma_{1}\sigma_{2}R(\mathfrak{r}_{1}-\mathfrak{r}_{4}+
\mathfrak{r}_{2}-\mathfrak{r}_{3}),\\
\mathcal{G}&= z\Gamma+ \sigma_{1}(R^{2}-\epsilon_{2})(\mathfrak{r}_{3}-\mathfrak{r}_{4})(\mathfrak{r}_{1}+\mathfrak{r}_{2}+R) +
\sigma_{2}(R^{2}+\epsilon_{2})(\mathfrak{r}_{1}-\mathfrak{r}_{2})(\mathfrak{r}_{3}+\mathfrak{r}_{4}-R)\\
&-2\sigma_{1}\sigma_{2} \left\{2R[\mathfrak{r}_{1}\mathfrak{r}_{2}-\mathfrak{r}_{3}\mathfrak{r}_{4}-\sigma_{1}(\mathfrak{r}_{1}-
\mathfrak{r}_{2})+\sigma_{2}
(\mathfrak{r}_{3}-\mathfrak{r}_{4})]-\epsilon_{2}(\mathfrak{r}_{1}-\mathfrak{r}_{4}+\mathfrak{r}_{2}-
\mathfrak{r}_{3})\right\},\\
\mathfrak{r}_{i}&:=a_{i}r_{i},\qquad a_{1}=\frac{s_{1+}}{\bar{s}_{1+}}, \qquad a_{2}=\frac{s_{1-}}{\bar{s}_{1-}}, \qquad a_{3}=\frac{s_{2-}}{\bar{s}_{2-}}, \qquad  a_{4}=\frac{s_{2+}}{\bar{s}_{2+}}, \qquad |a_{j}|\equiv 1, \\
\kappa_{o}&=\frac{64M^{3}\sigma_{1}\sigma_{2}(R^{4}-2\epsilon_{1}R^{2}+\epsilon_{2}^{2})
\left\{\mathfrak{q}^{2}\left[M(R^{2}+M^{2}-
2\Delta)+2\left(\mathfrak{q}(P_{1}+P_{2})+M\epsilon_{1}-R\epsilon_{2}\right)\right]-M(2P_{2}+M \mathfrak{q})^{2} \right\}}{\bar{s}_{1+}\bar{s}_{1-}\bar{s}_{2+}\bar{s}_{2-}},\\
s_{1\pm}&=\mathfrak{q}(P_{1}+P_{2})-M(\Delta + M R)-(R+M)(\epsilon_{2}\pm 2M\sigma_{1}) + i M [2P_{2}- \mathfrak{q}(R \pm 2\sigma_{1})], \quad r_{1,2}=\sqrt{\rho^{2}+(z-R/2 \mp \sigma_{1})^{2}},\\
s_{2\pm}&=\mathfrak{q}(P_{1}+P_{2})-M(\Delta - M R)-(R-M)(\epsilon_{2}\pm 2M\sigma_{2}) + i M [2P_{2}+\mathfrak{q}(R \pm 2\sigma_{2})], \quad r_{3,4}=\sqrt{\rho^{2}+(z+R/2 \mp \sigma_{2})^{2}}.
\label{7parameters}  \end{split} \eea
\end{widetext}

The above metric Eq.\ (\ref{7parameters}) contains seven parameters into the set $\{M,R,\mathfrak{q},\sigma_{1}, \sigma_{2},P_{1},P_{2}\}$, and it is reduced to the one that was already considered in Ref.\ \cite{MankoRuiz} in the absence of $J_{0}$. As a matter of fact, the advantage of this particular choice of the axis data Eq.\ (\ref{ernstaxis}) is that it gives us more information about several options to eliminate the NUT charge $J_{0}$ with the main purpose to obtain an asymptotically flat exact solution. It is worthwhile to mention, that contrary to was claimed by the authors of Ref.\ \cite{MankoRuiz}, the aforementioned metric Eq.\ (\ref{7parameters}) is not only exclusive to describe the interaction among two nonequal corotating Kerr BHs because it may also be used to define perfectly configurations of counterrotating BH systems, as we shall observe in Sec. III. As a matter of fact, the ansatz for solving the axis conditions is an extension of the one that has been used to describe unequal counter-rotating Kerr BHs \cite{ICM}, which is recovered after settling $\mathfrak{q}=0$ in Eqs.\ (\ref{ernstaxis}) and (\ref{7parameters}).

\vspace{-0.5cm}
\subsection{Two unequal Kerr sources apart by a strut: solving the axis conditions}
\vspace{-0.5cm}
By construction the metric Eq.\ (\ref{7parameters}) contains an elementary flatness in the upper part of the symmetry axis; it means that the conditions: $\omega(\rho=0,\alpha_{1}< z <\infty)=0$, and $\gamma(\rho=0,\alpha_{1}<z<\infty)=\gamma(\rho=0,-\infty<z<\alpha_{4})=0$ are automatically satisfied. Therefore, the remaining axis conditions on the symmetry axis are
\bea \begin{split}
\omega(\rho=0, \alpha_{2}<z< \alpha_{3})&=0, \\
\omega(\rho=0, -\infty<z< \alpha_{4})&=0,
\end{split}\label{omegasregions}\eea

\noi which in terms of the canonical parameters $\{\alpha_{n}, \beta_{i}\}$ are given by \cite{ICM}
\bea \begin{split} {\rm{Im}}\left[ \left|\begin{array}{ccccc}
0 &     1       &        1     &       1    &      1 \\
1 & \pm \gamma_{11}  & \pm \gamma_{12}  & \gamma_{13}& \gamma_{14}  \\
1 & \pm \gamma_{21} & \pm \gamma_{22} & \gamma_{23} & \gamma_{24} \\
0 & \kappa_{11} & \kappa_{12} & \kappa_{13} & \kappa_{14}\\
0 & \kappa_{21} & \kappa_{22} & \kappa_{23} & \kappa_{24}\\
\end{array}
\right| \right]=0,& \\
\gamma_{jn}=(\alpha_{n}-\beta_{j})^{-1}, \qquad \kappa_{jn}=(\alpha_{n}-\bar{\beta}_{j})^{-1}.&
\end{split}\label{axisconditions}\eea

The condition with $+$ sign is equivalent to kill the NUT charge (gravitomagnetic monopole), that is given explicitly above by Eq.\ (\ref{Multipolarterms}), while the other one containing a $-$ sign disconnects the region in between sources; it means that after solving such a condition, the mass in the middle region does not contribute to the total ADM mass \cite{ADM}, thus both sources will be apart by a massless strut. Due to the fact that we have at hand several possibilities among the seven parameters that could eliminate $J_{0}$, without taking into account the trivial case $\mathfrak{q}=0$, we are going to select the option
\be \epsilon_{1}= \frac{\Delta}{2}+ \frac{[\mathfrak{q}(P_{1}+P_{2})-\epsilon_{2} R]^{2}-M^{2}(4P_{1}P_{2}+\epsilon_{2}^{2})}{2M^{2}(R^{2}-\Delta)}.\label{factorizes}\ee

Surprisingly, the extremely complicated axis condition (with $-$ sign) from Eq.\ (\ref{axisconditions})  eventually is reduced to a quadratic equation for $\epsilon_{2}$, $P_{1}$ and $P_{2}$, namely
\begin{widetext}
\bea \begin{split}
 &\mathfrak{q}(R+M)\left[(R+M)(R^{2}-\Delta)-M\mathfrak{q}^{2}\right]\epsilon_{2}^{2}+
\left[M(R^{2}-\Delta)^{2}+2(R+M)\left(\Delta(R^{2}-\Delta)-M\mathfrak{q}^{2}R\right)\right]
(P_{1}+P_{2})\epsilon_{2}\\
&+\mathfrak{q}\left[M^{2}(R^{2}+MR+\mathfrak{q}^{2})(P_{1}-P_{2})^{2}
+(\Delta+MR)(\Delta-MR-R^{2})(P_{1}+P_{2})^{2}\right]-M^{2}(R^{2}-\Delta)\\
&\times \left\{\left[M\mathfrak{q}^{2}+(R+M)(R^{2}+MR+\mathfrak{q}^{2})\right](P_{1}-P_{2})-
M\mathfrak{q}(R+M)(R^{2}-\Delta)\right\}=0.
\label{conditionmiddle}\end{split}\eea

This quadratic equation is solved by adopting the following parametrization:
\bea \begin{split}
\epsilon_{2}&=- \frac{(\Delta+MR)(P_{1}+P_{2}) +M r(R^{2}-\Delta)}{\mathfrak{q}(R+M)},\\
P_{1,2}&=\frac{M\mathfrak{q}^{2}-(R+M)(R^{2}-\Delta)}{2{\left[(R+M)^{2}+\mathfrak{q}^{2}\right]}}
\,r-\frac{\mathfrak{q}^{2}(R^{2}+MR+\mathfrak{q}^{2})}{2(R+M)
\left[(R+M)^{2}+\mathfrak{q}^{2}\right]}\,\frac{s^{2}}{r} \pm \frac{R^{2}-\Delta}{2(R+M)}\,s \mp\frac{\mathfrak{q}(R^{2}-\Delta)}{2(R+M)}\\
&+\frac{\mathfrak{q}(R^{2}+MR+2\mathfrak{q}^{2})}{2(R+M)}\,\frac{s}{r}-\frac{\mathfrak{q}^{2}
\left[(R+M)^{2}+\mathfrak{q}^{2}\right]}{2(R+M)}\,\frac{1}{r},
\label{solution}\end{split}\eea
\end{widetext}

\noi where it is observed that there exists a symmetry property in our ansatz that solves the axis conditions since $P_{1,2}\rightarrow -P_{2,1}$, $\epsilon_{2}\rightarrow -\epsilon_{2}$, and $\epsilon_{1}\rightarrow\epsilon_{1}$, under the transformations $s\rightarrow s$,\, $r\rightarrow -r$. This special characteristic means that we are interchanging the location of the components of the BS as well as their physical properties. The solving of the axis conditions clearly illustrates that the affirmation made recently in Ref.\ \cite{MankoRuiz} on the need of very powerful computers to perform the required calculations in the analytical form is not correct.

\section{Physical representation for the black hole horizons}
\vspace{-0.5cm}
In order to obtain a real physical representation of the double-Kerr solution we must calculate the Komar parameters of the BS. To perform such a task we will use the well-known Tomimatsu formulas \cite{Tomimatsu0} for stationary axisymmetric spacetimes in vacuum
\bea \begin{split}
M_{i}&=-\frac{1}{8\pi}\int_{H_{i}} \omega\, {\rm{Im}}({\cal{E}}_{z}) d\varphi dz, \\ J_{i}&=-\frac{1}{8\pi}\int_{H_{i}}\omega\, \left(1+\frac{1}{2}\omega\, {\rm{Im}}({\cal{E}}_{z}) \right) d\varphi dz,\label{Tomi}\end{split}\eea

\noi where the integrals are evaluated over the BH horizons, which are defined as null hypersurfaces $H_{i}=\{\alpha_{2i}\leq z \leq \alpha_{2i-1}, \varphi \leq 2\pi,\, \rho\rightarrow 0\},$ \, $i=1,2$. Substituting Eqs.\ (\ref{7parameters}), (\ref{factorizes}), and (\ref{solution}) inside Eq.\ (\ref{Tomi}), one obtains the individual mass and angular momentum for each BH
\bea \begin{split}
M_{1,2}&=\frac{M}{2}\pm \frac{\mathfrak{q}[(R+M)^{2}+\mathfrak{q}^{2}]-(R^{2}+MR+\mathfrak{q}^{2})s}{2r(R+M)}, \\
J_{1,2}&=M_{1,2}\left(\frac{s \pm r}{2}\right).\label{massesmomenta}\end{split}\eea

\noi It follows that the total mass $M=M_{1}+M_{2}$ and total angular momentum $J=J_{1}+ J_{2}$, where
$s=a_{1}+a_{2}$ and $r=a_{1}-a_{2}$, being $a_{i}\equiv J_{i}/M_{i}$ the individual angular momentum per unit mass. On the other hand, from Eq.\ (\ref{Multipolarterms}) the following relation arises
\be J-M\mathfrak{q}=\frac{(R^{2}-\Delta)(\mathfrak{q}-a_{1}-a_{2})}{2(R+M)}, \label{Momentum}\ee

\noi and it is reduced to a dynamical law for interacting Kerr sources with struts via the expressions contained in Eq.\ (\ref{massesmomenta}), namely
\begin{widetext}
\bea \begin{split} &\mathfrak{q}^{3}-(a_{1}+a_{2})\mathfrak{q}^{2}+(R+M_{1}+M_{2})^{2}\mathfrak{q}
-(R+M_{1}+M_{2})[a_{1}(R+M_{1}-M_{2})+a_{2}(R-M_{1}+M_{2})].\label{condition}\end{split}\eea

Finally, combining Eqs.\ (\ref{factorizes}), (\ref{solution}), and (\ref{condition}), the explicit formula for both unequal $\sigma_{i}$ in terms of Komar physical parameters are given by
\bea \begin{split}
\sigma_{1}&=\sqrt{M_{1}^{2}-a_{1}^{2}+4a_{1}M_{2}\frac{a_{1}M_{2}\mathfrak{q}^{2}+
[M_{1}(\mathfrak{q}+a_{1}-a_{2})+a_{1}R][(R+M)^{2}+\mathfrak{q}^{2}]}
{[(R+M)^{2}+\mathfrak{q}^{2}]^{2}}},\\
\sigma_{2}&=\sqrt{M_{2}^{2}-a_{2}^{2}+4a_{2}M_{1}\frac{a_{2}M_{1}\mathfrak{q}^{2}+
[M_{2}(\mathfrak{q}-a_{1}+a_{2})+a_{2}R][(R+M)^{2}+\mathfrak{q}^{2}]}
{[(R+M)^{2}+\mathfrak{q}^{2}]^{2}}},
 \label{sigmas}\end{split}\eea

\noi where $\sigma_{2}=\sigma_{1 (1\leftrightarrow2)}$. The solving of the axis conditions Eq.\ (\ref{axisconditions}) and the physical functional form of each half-length horizon $\sigma_{i}$ are two of the principal results of this paper. It is worth mentioning that both horizons can be entirely depicted in terms of the five parameters $\{M_{1},M_{2},a_{1},a_{2},R\}$ after solving analytically the above cubic Eq.\ (\ref{condition}), whose roots explicitly are
\bea \begin{split}
\mathfrak{q}_{(k)}&= -\mathfrak{a}_{1} + e^{i 2\pi k/3}\left[\mathfrak{b}_{o}+ \sqrt{\mathfrak{b}_{o}^{2}-\mathfrak{a}_{o}^{3}}\right]^{1/3}+ e^{-i 2\pi k/3}\mathfrak{a}_{o} \left[\mathfrak{b}_{o}+ \sqrt{b_{o}^{2}-a_{o}^{3}}\right]^{-1/3},\\
\mathfrak{a}_{o}&:=\mathfrak{a}_{1}^{2}-\mathfrak{a}_{2},\qquad
\mathfrak{b}_{o}:=(1/2)\left[3\mathfrak{a}_{1}\mathfrak{a}_{2}-\mathfrak{a}_{3}-2\mathfrak{a}_{1}^{3}\right], \qquad k=0,1,2,\\
\mathfrak{a}_{1}&:=-(a_{1}+a_{2})/3, \quad \mathfrak{a}_{2}:=(R+M_{1}+M_{2})^{2}/3, \quad \mathfrak{a}_{3}:= -(R+M_{1}+M_{2})[a_{1}(R+M_{1}-M_{2})+a_{2}(R-M_{1}+M_{2})].
\label{theq} \end{split}\eea
\end{widetext}

\begin{figure}[ht]
\centering
\includegraphics[width=8.5cm,height=5.0cm]{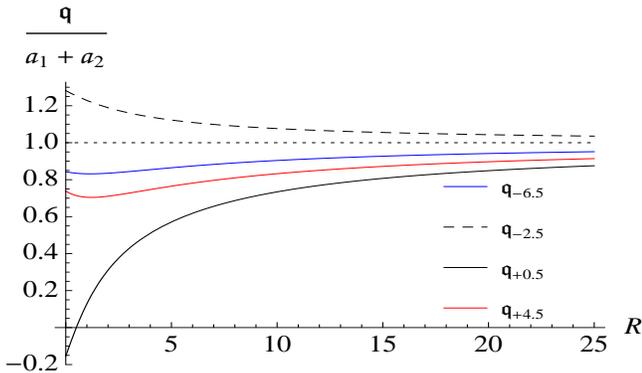}\\
\caption{Several figures of $\mathfrak{q}$ in the unequal case, for $M_{1}=1$ $M_{2}=2$, $a_{1}=1.5$, and different angular momentum values per unit mass $a_{2}$ indicated by the subindex.}
\label{theqnonextrem}\end{figure}

The parameter $\mathfrak{q}$ is the key for a better understanding of the dynamical interaction between two Kerr sources, and since this dynamical law is represented by a cubic equation, there exists at least one real root which in this case is given by the phase $k=0$. So, we have that $\mathfrak{q}$ can take positive or negative values depending on whether the configuration is co or counter-rotating as shown in Fig.\ \ref{theqnonextrem}. The constant line $\mathfrak{q}=0$ gives us the following condition among two nonequal counter-rotating Kerr BHs \cite{ICM}:
\be J_{2}=-\frac{J_{1}M_{2}}{M_{1}}\left(\frac{R+M_{1}-M_{2}}{R-M_{1}+M_{2}}\right),\label{thecounter}\ee

\vspace{-0.5cm}

\noi whose identical case $M_{1}=M_{2}=m$ and $a_{1}=-a_{2}=a$, represents one of the most simple exact solutions in which there is no need to solve the axis conditions. It was derived first in Ref.\ \cite{Varzugin,MRRS,Costa}. Moreover, after the redefinition $M_{1}=M_{2}=m$, $a_{1}=a_{2}=a\equiv j/m$, and $\mathfrak{q}= 2q$, from Eq.\ (\ref{condition}) one recovers the cubic equation for identical corotating Kerr sources \cite{CCLP}
\be  q^{3}- \frac{j}{m} q^{2}+ \frac{(R+2m)^{2}}{4}q- \frac{R(R+2m)j}{4m}=0. \label{theqidentical}\ee

\vspace{-0.5cm}
\subsection{Physical and thermodynamical properties}
\vspace{-0.5cm}
The first physical property we will consider in this two-body configuration is the interaction force associated with the strut; a line source of pressure deforming the BH horizons. It can be computed by means of the formula \cite{Israel,Weinstein}
\be \mathcal{F}=\frac{1}{4}(e^{-\gamma_{s}}-1),
\label{force1}\ee

\noi being $\gamma_{s}$ the metric function $\gamma$ evaluated on the middle region corresponding to the conical singularity among sources. It is really amazing how simple turns out to be the formula for the interaction force between two Kerr sources, which takes the final form
\be \mathcal{F}=\frac{M_{1}M_{2}[(R+M)^{2}-\mathfrak{q}^{2}]}
{(R^{2}-M^{2}+\mathfrak{q}^{2})[(R+M)^{2}+\mathfrak{q}^{2}]}.
\label{force}\ee

If $\mathfrak{q}=0$ and there is no rotation ($a_{i}=0$) we recover the first known expression of the force for two Schwarzschild BHs \cite{Bach}
\be \mathcal{F}=\frac{M_{1}M_{2}}{R^{2}-(M_{1}+M_{2})^{2}}.
\label{forceSCH}\ee

As a curious fact, when $\mathfrak{q}=0$, the force contains the same aspect like the above Eq.\ (\ref{forceSCH}) concerning to static BHs, and this is plausible whenever the five physical parameters satisfy the relation Eq.\ (\ref{thecounter}), that can be rewritten in the form
\be J_{1} + J_{2}+ R\left(\frac{J_{1}}{M_{1}}+ \frac{J_{2}}{M_{2}} \right)- M_{1}M_{2}
\left(\frac{J_{1}}{M_{1}^{2}}+\frac{J_{2}}{M_{2}^{2}}\right)=0. \label{vacuumrelation}\ee

The force tends to zero as the sources move further and further away from each other. In this case if $R \rightarrow \infty$, the parameter $\mathfrak{q}\rightarrow  a_{1}+a_{2}$, and the force contains the following aspect:
\begin{widetext}
\be \mathcal{F}\simeq\frac{M_{1}M_{2}}{R^{2}}\left[1+\frac{(M_{1}+M_{2})^{2}-3(a_{1}+a_{2})^{2}}
{R^{2}}+ \frac{4(a_{1}+a_{2})[M_{1}a_{1}+M_{2}a_{2}+4(M_{2}a_{1}+M_{1}a_{2})]}{R^{3}}+ O\left(\frac{1}{R^{4}}\right)\right],\ee
%\end{widetext}

\noi which is the formula already given by Dietz and Hoenselaers \cite{DH} with an extra term containing more information about the spin-spin interaction at large distances. It should be remarked that in the limit $R \rightarrow \infty$ is also recovered from Eq.\ (\ref{sigmas}) the expression $\sigma_{i}=\sqrt{M_{i}^{2}-J_{i}^{2}/M_{i}^{2}}$ for one isolated Kerr BH.
\end{widetext}

%\vspace{-0.5cm}
\begin{figure}[ht]
\centering
\includegraphics[width=8.5cm,height=5.0cm]{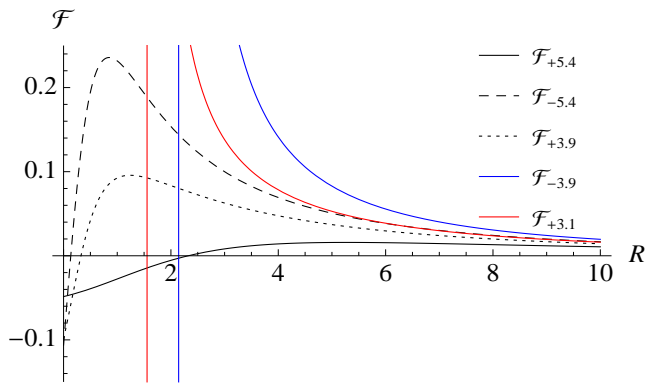}\\
\caption{The behaviour of the interaction force for $M_{1}=1$, $M_{2}=2$, $a_{1}=1.5$, and several angular momentum values per unit mass $a_{2}$ labeled with a subindex. The merging limit is indicated by
a vertical asymptote. }
\label{theforce}\end{figure}

Turning now our attention to the thermodynamical features of the binary system, where is well-known that each BH fulfills the Smarr formula \cite{Smarr}
\be M_{i}=\frac{\kappa_{i}S_{i}}{4\pi}+2\Omega_{i}J_{i}=\sigma_{i}+2\Omega_{i}J_{i}, \qquad i=1,2,\ee

\noi being $\kappa_{i}$ the so-called surface gravity of the $i \rm {th}$ BH, which is related to the corresponding angular velocity $\Omega_{i}$ by means of \cite{DH,Tomimatsu}
\be \kappa_{i}= \sqrt{-\Omega_{i}^{2}e^{-2\gamma^{H_{i}}}}, \qquad \Omega_{i}:= \omega_{i}^{-1},\ee

\noi where $\omega_{i}$ and $\gamma^{H_{i}}$ are the constant values of the metric functions $\omega$ and $\gamma$ on the axis part associated to each horizon $H_{i}$. Additionally, $S_{i}$ is the area of the horizon. Taking into account Eqs.\ (\ref{condition}) and (\ref{sigmas}) it follows that the angular velocities, surface gravities and the area of the horizons acquire the final compact expressions
\bea \begin{split} \Omega_{1}&= \frac{M_{1}-\sigma_{1}}{2J_{1}}= \frac{J_{1}F_{1}}{2M_{1}^{2}(M_{1}+\sigma_{1})}, \\
\kappa_{1}&= \frac{\sigma_{1}P_{0}[(R+\sigma_{1})^{2}
-\sigma_{2}^{2}]}{[P_{0}(M_{1}+\sigma_{1})-2M_{1}a_{1}\mathfrak{q}]^{2}+a_{1}^{2}(R^{2}-\Delta)^{2}},\\
S_{1}&=4\pi \frac{[P_{0}(M_{1}+\sigma_{1})-2M_{1}a_{1}\mathfrak{q}]^{2}+a_{1}^{2}(R^{2}-\Delta)^{2}}{P_{0}[(R+\sigma_{1})^{2}
-\sigma_{2}^{2}]},\\
F_{1}&:=1-\frac{4M_{2}}{a_{1}}\left[\frac{a_{1}M_{2}\mathfrak{q}^{2}+
[M_{1}(\mathfrak{q}+a_{1}-a_{2})+a_{1}R]P_{0}}{P_{0}^{2}}\right],\\
P_{0}&:=(R+M)^{2}+\mathfrak{q}^{2},\\
\Omega_{2}&=\Omega_{1(1\leftrightarrow 2)},\quad \kappa_{2}=\kappa_{1(1\leftrightarrow 2)},
\quad S_{2}=S_{1(1\leftrightarrow 2)}.
\end{split}\eea

On the other hand, in order to interpret the interaction between BHs, by looking once more the denominator of the above formula of the force, at first sight it seems that the merging limit occurs whether $R$ tends to a minimal value given by $R_{0}=\sqrt{M^{2}-\mathfrak{q}^{2}}=\sigma_{1}+\sigma_{2}$, from which the interaction force $\mathcal{F}\rightarrow \infty$. At this particular value of the distance we notice from Eq.\ (\ref{condition}) [or Eq.\ (\ref{Momentum})] that $\mathfrak{q}=J/M$, and therefore the values of the half-length horizons are
\bea \begin{split}
\sigma_{1}&=\sqrt{M_{1}^{2}-\frac{a_{1}M_{1}^{2}[a_{1}(R_{0}+M_{1}-M_{2})+2M_{2}a_{2}]}{(M_{1}+M_{2})^{2}
(R_{0}+M_{1}+M_{2})}},\\
\sigma_{2}&=\sqrt{M_{2}^{2}-\frac{a_{2}M_{2}^{2}[a_{2}(R_{0}-M_{1}+M_{2})+2M_{1}a_{1}]}{(M_{1}+M_{2})^{2}
(R_{0}+M_{1}+M_{2})}}.
 \label{sigmasmeerginglimit}\end{split}\eea

Let us now consider that both BHs become extremal; i.e., $\sigma_{1}=\sigma_{2}=0$, thus the minimal value $R_{0}=0$ befalls when $\mathfrak{q}=M$, and because $\mathfrak{q}=J/M$, the MP will produce a single extreme BH of mass $M=M_{1}+M_{2}$ and total angular momentum $J=J_{1}+J_{2}$, satisfying a well-known relation given by
\be J_{1}+J_{2}=(M_{1}+M_{2})^{2}. \label{extremerelation} \ee

Then, a natural question arises: What values the angular momenta of the BHs are taking during the MP in this extreme case? The answer of this question comes immediately from Eq.\ (\ref{sigmasmeerginglimit}), by settling $\sigma_{1}=\sigma_{2}=0$; having that
\be J_{1}=M_{1}(M_{1}+M_{2}), \qquad J_{2}=M_{2}(M_{1}+M_{2}),\ee

\noi whose sum $J_{1}+J_{2}$ clearly recovers Eq.\ (\ref{extremerelation}). Apparently, this description of the MP corresponds to a BS composed by corotating sources; in agreement with Ref.\ \cite{Costa} in the identical case. Moreover, if $ \mathfrak{q}=0$, and therefore $R=M_{1}+M_{2}$, the BH horizons become statics during the merging limit since $\sigma_{1}=M_{1}$ and $\sigma_{2}=M_{2}$ \cite{ICM}, as well as identical when both sources result to be extreme \cite{Costa,ICM}. In both static cases the merging limit begins to form a single Schwarzschild BH containing a total mass equal to the sum $M_{1}+M_{2}$ whose area of its horizon satisfies the relation $S=S_{1}+S_{2}=16 \pi (M_{1}+M_{2})^{2}$.

Continuing with the analysis, after assigning positive values for both masses, the balance condition and the absence of the strut ($\mathcal{F}=0$) is not fulfilled before the MP happens; i.e., the force is never crossing the horizontal axis if both masses are positives. This statement was confirmed by choosing a wide range of numerical values and it agrees with Ref.\ \cite{DH}. Additionally, Fig.\ \ref{theforce} shows several shapes of the force which can be attractive/repulsive taking positive/negative values before or after the MP is occurring, and can even acquire the value of zero. Regarding the last point, in the search of equilibrium states in which the gravitational attraction is counterbalancing the spin-spin interaction, we observe from Eq.\ (\ref{force}) that the strut disappears with the condition $\mathfrak{q}=-\varepsilon(R+M)$, \, $\varepsilon=\pm 1$, and it is leading us to an equilibrium law that has been studied in several papers by Manko and coauthors. We will refer to the last one research with the purpose to derive once again the final formulas for $\sigma_{i}$ obtainable from Eq.\ (\ref{sigmas}) in this equilibrium situation \cite{MR}
\begin{widetext}
\bea \begin{split}
\sigma_{1}&=\sqrt{M_{1}^{2}-a_{1}^{2}+M_{2}a_{1}\frac{a_{1}(M+M_{1}+2R)-2M_{1}
[a_{2}+\varepsilon(M+R)]}{(M+R)^{2}}},\\
\sigma_{2}&=\sqrt{M_{2}^{2}-a_{2}^{2}+M_{1}a_{2}\frac{a_{2}(M+M_{2}+2R)-2M_{2}
[a_{1}+\varepsilon(M+R)]}{(M+R)^{2}}},
 \label{sigmasequilibrium}\end{split}\eea
\end{widetext}

\noi whereas the equilibrium law is derived from Eq.\ (\ref{condition}); it reads \cite{MR}
\be J_{1}+J_{2}+R\left(\frac{J_{1}}{M_{1}}+\frac{J_{2}}{M_{2}}\right)+\varepsilon(R+M_{1}+M_{2})^{2}=0. \ee

\vspace{-0.5cm}
\begin{table}[ht]
\centering
\caption{Numerical values for equilibrium states during the MP, fixing the values in the masses $M_{1}=1$ and $M_{2}=2$.  }
\begin{tabular}{c c c c c c}
\hline \hline
$\sigma_{1}$&$\sigma_{2}$&$ a_{1}$ & $a_{2}$ & $R$ & $\mathfrak{q}$ \\ \hline
  1.36892 & 0.96607 & 1.2  & -5.2 & 0.09545  & -3.09545  \\
  0.78173 & 0.39310 & 0.7  & 4.2 & 0.08443 & 3.08443   \\
  1.62221 & 1.29883 & -2.0  & 5.6 & 0.08062 & 3.08062 \\
  1.40981 & 0.29261 & -1.44  & 5.42 & 0.18168 & 3.18168  \\
  2.36235 & 1.27626 & 5.0  & -7.4 & 0.20998 & -3.20998  \\
  \hline \hline
\end{tabular}
\label{table1}
\end{table}

It is well-known that two rotating BHs cannot be at equilibrium without a supporting strut before the system is merging since at least one of the sources develops ring singularities off the axis, due mainly to the presence of negative masses in the DKN solution \cite{DH,MRS,Hennig}. Nevertheless, after the MP befalls there exist equilibrium states for which both sources contain positive masses and can be observed as subextreme sources; i.e., BHs. For instance, in Fig.\ \ref{theforce} the curve labeled with the subindex given by the value $a_{2}=-5.4$ establishes an equilibrium state at $R\simeq 0.13247$. As far as we know this surprising phenomenon cannot be observed if the two sources carry equal masses and equal angular momenta in co and counter-rotating BS \cite{Varzugin,MRRS,Costa,CCLP}. So, apparently the aforementioned formula of the interaction force Eq.\ (\ref{force}) for nonequal spinning bodies can reveal more information on the possibility of finding equilibrium states during the collision of two BHs, but preserving positive both masses. Table \ref{table1} shows a set of numerical values satisfying equilibrium states during the MP, where it is observed that the condition contained in the Smarr formula $M_{i}>\sigma_{i}$ is not satisfied for some cases; in particular for the mass with value $M_{1}=1$. This physical aspect can be understood in a naive sense like the BH is forced to increase its ergoregion to compensate for its loss of rotation.

\vspace{-0.5cm}
\section{Final remarks}
\vspace{-0.5cm}
This paper is devoted to conclude one of the main problems during almost the last four decades that might help to study in the most general case dynamical and thermodynamical aspects of two interacting BHs in stationary axisymmetric vacuum systems: \emph{the solving of the axis conditions}. Our suitable parametrization of the double Kerr solution \cite{KramerNeugebauer} including the NUT charge led us to consider the desirable parametrization which eventually simplified and helped us to solve the axis condition in between sources. After that, we have been capable to obtain the nontrivial expressions for the BH horizons $\sigma_{i}$, \, $i=1,2$, in terms of the five arbitrary physical Komar parameters $\{M_{1},M_{2},a_{1},a_{2},R\}$ as well as the thermodynamical features included in the Smarr formula \cite{Smarr}. These five physical parameters are contained within the coefficients of a cubic equation which is interpreted as a dynamical law for interacting BHs, and it reduces to the equilibrium law for two arbitrary Kerr sources \cite{MR} in the absence of the strut. Another interesting physical property of the BS reveals that there exist equilibrium states without a strut during the MP on which the sources can be noticed as subextreme carrying both positive masses; this point is certainly intriguing an we hope to delve into it in greater depth in the future. It is worthwhile to mention, that the physical representation considered in this work is more transparent at the moment of considering astrophysical phenomena related to GW, like the collision of two BHs since the quasinormal modes of GW will be in terms of these physical parameters. We are convinced that our results will help not only to derive further exact models regarding geodesics around binary BHs with the main objective to research GW during the MP, but also they could be useful at the moment of studying their shadows and lensing effects like those considered earlier in Ref. \cite{Cunha}.

We end up this section by mentioning that the path used in this paper on the resolution of the axis conditions can be extended to the more complicated electrovacuum systems. In this case the axis data of the Ernst potentials will show the following aspect:
\begin{widetext}
\bea \begin{split}
{\cal E}(0,z)&=\frac{e_{1}}{e_{2}}, \qquad \Phi(0,z)=\frac{(\mathcal{Q}+i\mathcal{B})z+q_{o}+i b_{o}}{e_{2}}, \\
e_{1}&=z^{2}-[M + i(\mathfrak{q}+2J_{0})]z + \frac{M(2\Delta_{o}-R^{2})-2(M\epsilon_{1}+R\epsilon_{2})-4(\mathcal{Q} q_{o}+ \mathcal{B} b_{o})}{4M} + \frac{\mathfrak{q}(P_{1} +P_{2})}{2M} -2 \mathfrak{q} J_{0}\\
&+ i\left(P_{1}- \frac{2J_{0}(P_{2}+M \mathfrak{q})}{\mathfrak{q}} \right), \\
e_{2}&=z^{2} + (M -i\mathfrak{q})z + \frac{M(2\Delta-R^{2})-2(M\epsilon_{1}-R\epsilon_{2})+4(\mathcal{Q} q_{o}+ \mathcal{B} b_{o})}{4M} -\frac{\mathfrak{q}(P_{1} +P_{2})}{2M} + i P_{2}, \\
J_{0}&= \frac{\mathfrak{q}}{2M} \left( \frac{[\mathfrak{q}(P_{1}+P_{2})-\epsilon_{2}R-2(\mathcal{Q} q_{o}+ \mathcal{B} b_{o})]^{2}-
M^{2}\left[4(P_{1}P_{2}+q_{o}^{2}+b_{o}^{2})+
(R^{2}-\Delta_{o})(2\epsilon_{1}-\Delta_{o})+\epsilon_{2}^{2}\right]}{\mathfrak{q}^{2}\left[M(R^{2}+M^{2}-
2\Delta_{o})+2\left(\mathfrak{q}(P_{1}+P_{2})+M\epsilon_{1}-R\epsilon_{2}-2(\mathcal{Q} q_{o}+ \mathcal{B} b_{o})\right)\right]-M(2P_{2}+M \mathfrak{q})^{2}} \right), \\
\Delta_{o}&:= M^{2}-\mathfrak{q}^{2}-\mathcal{Q}^{2}-\mathcal{B}^{2}, \label{ernstaxiselectro}\end{split}\eea
\end{widetext}

\noi where $\mathcal{Q}$ and $\mathcal{B}$ are the total electric and magnetic charges, respectively. Furthermore, $q_{o}$ and $b_{o}$ can be associated (but they are not equally!) to the electric and magnetic dipole moments. Therefore, if one could be interested in the study of configurations of nonequal interacting Kerr-Newman BHs, it is needed to solve the axis conditions for electrovacuum systems in combination with the condition that eliminates the two individual magnetic charges on each BH source \cite{Tomimatsu}, namely
\be A_{4}(\rho=0,z=\alpha_{2i-1})-A_{4}(\rho=0,z=\alpha_{2i})=0, \quad
i=1,2. \ee

This procedure was employed in Ref.\ \cite{RICM} by using the aforementioned axis data Eq.\ (\ref{ernstaxiselectro}) with $\mathfrak{q}=0$ and $\mathcal{B}=0$, where the seven physical parameters of the Kerr-Newman BS establish a dynamic scenario through the relation
\begin{widetext}
\bea \begin{split}
&M_{1}M_{2}(R+M_{1}+M_{2})\left[ J_{1}+J_{2}+R\left(\frac{J_{1}}{M_{1}}+\frac{J_{2}}{M_{2}}\right)-M_{1}M_{2}
\left(\frac{J_{1}}{M_{1}^{2}} +\frac{J_{2}}{M_{2}^{2}}\right)\right]\\
&+ (M_{1}-M_{2})(Q_{1}+Q_{2})(Q_{1}J_{2}-Q_{2}J_{1})-Q_{1}Q_{2}(J_{1}+J_{2})R=0, \end{split}\eea
\end{widetext}

\noi which generalizes Eq.\ (\ref{vacuumrelation}) since now the electric charges $Q_{1}$ and $Q_{2}$ are included. We expect to accomplish in a future the description of the most general physical problem regarding the interaction of two Kerr-Newman BHs separated by a massless strut.

%\vspace{-0.5cm}
\section*{ACKNOWLEDGMENTS}
\vspace{-0.5cm}
This work was supported by PRODEP, M\'exico, grant no 511-6/17-7605 (UACJ-PTC-367).

\end{document}